\documentclass[12pt,preprint]{aastex}
\newcounter{address}
\newcommand{\latin}[1]{\textit{#1}}

\newcommand{\eg}{\latin{e.g.}}
\newcommand{\ie}{\latin{i.e.}}
\newcommand{\Ie}{\latin{I.e.}}
\newcommand{\vs}{\latin{vs.}}

\newcommand{\Hipparcos}{\textit{Hipparcos}}
\newcommand{\normal}{{\cal N}}
\renewcommand{\vec}[1]{\mathbf{#1}} 
\newcommand{\inv}{^{-1}}
\newcommand{\bb}{\vec{b}}
\newcommand{\cc}{\vec{c}}

\newcommand{\mm}{\vec{m}}
\newcommand{\vv}{\vec{v}}
\newcommand{\ww}{\vec{w}}
\newcommand{\bij}{\bb_{ij}}
\newcommand{\bbij}{\bij}
\newcommand{\cci}{\cc_i}
\newcommand{\eex}{\vec{\hat{x}}}
\newcommand{\eey}{\vec{\hat{y}}}
\newcommand{\eez}{\vec{\hat{z}}}

\newcommand{\eel}{\vec{\hat{l}}_i}
\newcommand{\eeb}{\vec{\hat{b}}_i}
\newcommand{\mmj}{\mm_j}
\newcommand{\mmk}{\mm_k}
\newcommand{\vvi}{\vv_i}

\newcommand{\vvdisk}{\vv_\mathrm{disk}}
\newcommand{\Kdisk}{K_\mathrm{disk}}
\newcommand{\vvhalo}{\vv_\mathrm{halo}}
\newcommand{\vvlsr}{\vv_\mathrm{LSR}}
\newcommand{\vvsun}{\vv_\odot}
\newcommand{\wwi}{\ww_i}

\newcommand{\ten}[1]{\mathbf{#1}} 
\newcommand{\BB}{\ten{B}}
\newcommand{\CC}{\ten{C}}
\newcommand{\QQ}{\ten{Q}}
\newcommand{\RR}{\ten{R}}
\renewcommand{\SS}{\ten{S}}
\newcommand{\TT}{\ten{T}}
\newcommand{\VV}{\ten{V}}
\newcommand{\BBij}{\BB_{ij}}
\newcommand{\CCi}{\CC_i}
\newcommand{\QQi}{\QQ_i}
\newcommand{\RRi}{\RR_i}
\newcommand{\SSi}{\SS_i}
\newcommand{\VVj}{\VV_{\!j}} 
\newcommand{\VVdisk}{\VV_\mathrm{\!disk}}
\newcommand{\VVhalo}{\VV_\mathrm{\!halo}}
\newcommand{\TTij}{\TT_{ij}}

\newcommand{\T}{^{\scriptscriptstyle \top}}   
\newcommand{\tr}{\mathrm{tr}}                 
\newcommand{\alphaj}{\alpha_j}
\newcommand{\alphak}{\alpha_k}

\newcommand{\alphahalo}{\alpha_\mathrm{halo}}
\newcommand{\qij}{q_{ij}}
\newcommand{\qqj}{q_j}
\newcommand{\subsamplecolor}{ 0.654<(B-V)< 0.685}

\begin{document}

\title{
  Modeling complete distributions with incomplete observations:\\
  The velocity ellipsoid from \Hipparcos\ data.
}
\author{
  David~W.~Hogg\altaffilmark{\ref{NYU},\ref{email}},
  Michael~R.~Blanton\altaffilmark{\ref{NYU}},
  Sam~T.~Roweis\altaffilmark{\ref{Toronto}} and
  Kathryn~V.~Johnston\altaffilmark{\ref{Wesleyan}}
}

\setcounter{address}{1}
\altaffiltext{\theaddress}{\stepcounter{address}\label{NYU}
Center for Cosmology and Particle Physics, Department of Physics, New
York University, 4 Washington Place, New York, NY 10003}
\altaffiltext{\theaddress}{\stepcounter{address}\label{email}
\texttt{david.hogg@nyu.edu}}
\altaffiltext{\theaddress}{\stepcounter{address}\label{Toronto}
Department of Computer Science, University of Toronto}
\altaffiltext{\theaddress}{\stepcounter{address}\label{Wesleyan}
Department of Astronomy, Wesleyan University}

\begin{abstract}
An algorithm is developed to model the three-dimensional velocity
distribution function of a sample of stars using only measurements of
each star's two-dimensional tangential velocity.  The algorithm works
with ``missing data'': it reconstructs the three-dimensional
distribution from data (velocity measurements) every one of which has
one dimension unmeasured (the radial direction).  It also accounts for
covariant measurement uncertainties on the tangential velocity
components.  The algorithm is applied to tangential velocities
measured in a kinematically unbiased sample of 11,865 stars taken from
the \Hipparcos\ catalog, chosen to lie on the main sequence and have
well-measured parallaxes.  The local stellar velocity distribution
function of each of a set of 20 color-selected subsamples is modeled
as a mixture of two three-dimensional Gaussian ellipsoids of arbitrary
relative responsibility.  In the fitting, one Gaussian (the ``halo'')
is fixed at the known mean velocity and velocity variance tensor of
the Galaxy halo, and the other (the ``disk'') is allowed to take
arbitrary mean and arbitrary variance tensor.  The mean and variance
tensor (commonly the ``velocity ellipsoid'') of the disk velocity
distribution are both found to be strong functions of stellar color,
with long-lived populations showing larger velocity dispersion, slower
mean rotation velocity, and smaller vertex deviation than short-lived
populations.  The local standard of rest (LSR) is inferred in the
usual way and the Sun's motion relative to the LSR is found to be
$(U,V,W)_{\odot}=(10.1,4.0,6.7)\pm(0.5,0.8,0.2)~\mathrm{km\,s^{-1}}$.
Artificial data sets are made and analyzed, with the same error
properties as the \Hipparcos\ data, to demonstrate that the analysis
is unbiased.  The results are shown to be insensitive to the
assumption that the velocity distributions are Gaussian.
\end{abstract}

\keywords{
  Galaxy: fundamental parameters
  ---
  Galaxy: kinematics and dynamics
  ---
  methods: statistical
  ---
  solar neighborhood
  ---
  stars: kinematics
}

\section{Introduction}

The classical picture of the evolution of the velocity structure in
the Galactic disk is that stars are born within low-dispersion
clusters from cool gas on near-circular orbits.  These clusters
evaporate and the stellar orbit distribution is heated through
gravitational perturbations to the smooth disk potential.  Over time a
stellar population's velocity dispersion grows and its mean motion
lags behind that of pure circular orbits at the same Galactocentric
radius.  Thus, the velocity distribution of stars in the Solar
Neighborhood has been characterized as an ellipsoid whose centroid,
size and orientation varies systematically with the lifetimes (and
hence colors) of the stars under investigation
\citep[\eg, ][]{dehnen98a}.

This field has undergone a recent renaissance with the release of the
\Hipparcos\ data set of proper motions and parallaxes, measured with
accuracies of a few milliarcseconds.  Studies using these data to
analyze the local velocity distribution of stars can be broadly split
into two categories:

First, there are determinations of the moments of the velocity
distribution as a function of color assuming (as above) that it can be
described by a mean velocity and a single velocity dispersion tensor
\citep{dehnen98a, bienayme99a}.  These have led to more stringent
limits on the Solar Motion relative to a (hypothetical)
zero-dispersion population (the Local Standard of Rest), the age of
the Galactic disk and rate of heating of stellar populations
\citep{binney00a}.

Secondly, there are non-parametric derivations of the full three
dimensional velocity distribution function \citep{dehnen98b,
skuljan99a, chereul98a}.  These have revealed that the velocity
distribution is poorly described by a single ellipsoid; in fact it
contains significant structures on smaller velocity scales.
Importantly, the structures do not seem to be dominated by short-lived
stars.  These structures can be variously interpreted as trails from
evaporating clusters \citep{chereul99a}, or overdensities induced by
resonances in the disk associated with the bar \citep{dehnen00a,
fux01a} and/or spiral arms \citep{quillen03a}.

Our long-term goal is to pursue the latter category of project; \ie,
to develop algorithms to locate, understand the significance of, and
characterize, nontrivial structures in the velocity distribution, not
just in the local Galaxy but in the Galaxy halo.  These projects will
require new space-based astrometry data (\eg, what we expect from the
upcoming GAIA mission) in combination with large ground-based surveys
(\eg, Sloan Digital Sky Survey, \citealt{york00a}; Two-Micron All-Sky
Survey, \citealt{skrutskie97a}; Grid Giant Star Survey,
\citealt{majewski00a}).  In the short term, we have begun by analyzing
the \Hipparcos\ data set with a very general algorithm for fitting
distribution functions to data measured with nontrivial error
covariances and missing information.

From a computer science or nonlinear statistics perspective, these
problems fall into the category of ``missing data'' problems, in which
one constructs a model of an object (here the velocity
distribution function) using data points (here tangential velocities)
every one of which is incomplete (because, in this case, it has no radial
information).  We present a framework for a large set of
algorithms for solving such problems, and the details of the specific
restriction to the velocity distribution function as measured with
velocities projected onto the sphere.

In this paper we further restrict our attention on the trial problem
of re-deriving the properties of the velocity ellipsoid near the Sun.
In what follows, it is assumed that any color-selected, kinematically
unbiased sample of stars has a velocity distribution function which
can be modeled (for the purposes of measuring its velocity variance)
by a sum of two Gaussan ellipsoids, one for ``halo stars'' and one for
``disk stars'', and later, by a sum or mixture of $K>2$ Gaussian
ellipsoids.  Model parameters are chosen to maximize the total
likelihood of the \Hipparcos\ measurements (which, in this case, are
two-dimensional tangential velocity vectors), given their
uncertainties (which are two-dimensional covariance tensors); \ie, the
results presented here represent the optimization of an explicit,
justified, scalar objective function.  Our work differs from previous
work in several respects: we have the scalar objective function, we
present tests of the algorithm with relatively realistic artificial
data, and we relax the Gaussian assumption (\ie, expand the space of
allowed distribution functions).

In later papers in this series, we intend to generalize our
parameterization (to multi-modal disk distributions), locate
velocity-space structures, measure their statistical significance, and
characterize their properties.  This phenomenology will be essential
for distinguishing the various pictures for the origin of the velocity
sub-structure in the Galaxy disk.

\section{Model and algorithm}

In what follows the standard Galactic velocity coordinate system is
used, with directions $x$, $y$, and $z$ (and associated unit vectors
$\eex$, $\eey$, and $\eez$) pointing towards the Galactic center,
pointing in the direction of circular orbital motion, and pointing
towards the north Galactic pole, respectively.  Vectors will be
implicitly defined to be column matrices, so $\vec{a}\T\,\vec{b}$ is
the scalar product and $\vec{a}\,\vec{b}\T$ is a rank-2 tensor.  The
components $\eex\T\,\vv$, $\eey\T\,\vv$, and $\eez\T\,\vv$ of a
velocity $\vv$ are conventionally named ``$U$'', ``$V$'', and ``$W$''.

We treat any color-selected population of stars from \Hipparcos\ as
being composed of two kinematically distinct populations of stars, a
``halo'' population with velocity distribution described by a Gaussian
ellipsoid in velocity space with a mean velocity $\vvhalo$ with
respect to the Sun and velocity dispersion (variance) tensor
$\VVhalo$, with these parameters fixed at
\begin{eqnarray}\displaystyle
\vvhalo &=& [-220~\mathrm{km\,s^{-1}}]\,\eey \nonumber \\
\VVhalo &=& [100~\mathrm{km\,s^{-1}}]^2\,
            [\eex\,\eex\T+\eey\,\eey\T+\eez\,\eez\T]
\label{eq:haloparameters}
\end{eqnarray}
\citep{sirko04a}, plus a ``disk'' population described by another
Gaussian ellipsoid with mean $\vvdisk$ and dispersion tensor
$\VVdisk$, both of which are allowed to vary arbitrarily.  The
relative amplitude $\alphahalo$ of the halo Gaussian (\ie, the
fraction of stars in the halo) is also allowed to vary arbitrarily.
Sensitivity of the results to the assumed halo velocity dispersion of
$100~\mathrm{km\,s^{-1}}$ is discussed below.

The vast majority ($\sim 99$~percent) of the sample is expected to be
members of the disk population.  However, the inclusion of a halo
Gaussian prevents halo stars from distorting the measurement of the
disk velocity variance.  In effect, the halo Gaussian ``clips out''
velocity outliers in a responsible way.

Almost all the difficulty in inferring the parameters of this model,
\ie, $\vvdisk$ (three parameters), $\VVdisk$ (six parameters), and the
relative responsibility of the halo Gaussian (one parameter), comes
from the fact that \Hipparcos\ does not measure the total three-space
velocity $\vv$ of each star, but only its two-dimensional tangential
projection.

\subsection{Model generalities}

The approach developed here is extremely general and can be applied to
many different density estimation tasks in the presence of partially
observed data.  The assumption is that there are low-dimension
observations $\wwi$, which are noisy projections of higher-dimension
``true values'' $\vvi$:
\begin{equation}
\wwi=\RRi\,\vvi + \mathrm{noise} \;\;\;,
\end{equation}
where the $\RRi$ are known, non-square (or zero-determinant) projection
matrices, and the noise is drawn from a Gaussian with zero mean and
known (low-dimension) covariance tensor $\SSi$.  It is also assumed
that the $\vvi$ are drawn independently and identically distributed
from a probability distribution function $p(\vv)$ in the
higher-dimension space. The goal is to fit a model for $p(\vv)$
using only the incomplete observations $\{\wwi\}$, their 
covariances $\{\SSi\}$ and the projection matrices $\{\RRi\}$.

Note that there is no assumption that all data points have similar
non-square projection matrices; in fact the projection matrices 
(and thus the observations) may have different dimensionalities.

The density model $p(\vv)$ is parameterized as a mixture of $K$
Gaussians:

\begin{eqnarray}\displaystyle
  p(\vv) &=& \sum_{j=1}^K \alphaj\,\normal(\vv|\mmj,\VVj) \;\;\;,
\end{eqnarray}
where the amplitudes or ``rates'' $\alphaj$ sum to unity
and the function $\normal(\vv|\mm,\VV)$ is the normal (Gaussian)
distribution with mean $\mm$ and variance tensor $\VV$.

For a known projection matrix $\RRi$ and noise covariance $\SSi$ in
$\ww$ space (the lower-dimensional space of the observations) each
component of the mixture marginalizes to a lower-dimensional Gaussian,
and so the induced density is a conditional mixture of Gaussians on
$\ww$:
\begin{eqnarray}\displaystyle
      p(\ww,\vv,j) &=& p(\ww|\vv)\,p(\vv|j)\,p(j)\nonumber \\
    p(\ww|\RR,\SS) &=& \sum_j \int_{\vv} p(\ww|\vv)\,p(\vv|j)\,p(j)\,
                       \mathrm{d}\vv\nonumber \\
p(\ww|\vv,\RR,\SS) &=& \normal(\ww|\RR\,\vv,\SS)\nonumber \\
 p(\wwi|\RRi,\SSi) &=& \sum_{j=1}^K \alphaj\,\normal(\wwi|\RRi\,\mmj,\TTij)\nonumber \\
             \TTij &=& \RRi\,\VVj\,\RRi\T+\SSi \;\;\;,
\end{eqnarray}
where functions like $p(x,y,z)$ are joint probability distribution
functions of $x$, $y$, and $z$, and functions like $p(x|y)$ are
probability distribution functions of $x$ given (or at a specific
value of) $y$.  All other symbols are described above, except $\TTij$,
which is the combined variance for each measurement $i$ under the
assumption that it is drawn from Gaussian $j$, with part of the
variance coming from the (projected) variance $\VVj$ of the Gaussian,
and part coming from the measurement uncertainty variance $\SSi$.

This model will be called the ``projected mixture of Gaussians'' model
hereafter.


The objective of the fitting procedure is to maximize the conditional
likelihood of the entire set of low-dimensional projected observations
given the nonsquare matrices and the error covariances. In particular,
we are fitting for the means $\{\mmj\}$, variance tensors $\{\VVj\}$
and amplitudes $\{\alpha_j\}$ of the mixture of Gaussians in the
high-dimensional (unobserved) space.  Assuming the noise on each
observation is independent of other observation noises this (log)
likelihood is
\begin{equation}
\phi = \sum_i \ln p(\wwi|\RRi,\SSi) = \sum_i \ln \sum_{j=1}^K \alphaj\,\normal(\wwi|\RRi\,\mmj,\TTij) \;\;\;.
\end{equation}

The model parameters can be optimized in several ways.  One approach
is to directly compute gradients and use a generic optmizer to ascend
the objective; this is complicated by restrictions on the parameters
(e.g. the variances must be symmetric and positive-definite, the
amplitudes must be non-negative and sum to unity). Another approach is
to view the high-dimensional quantities as hidden variables and use
the expectation-maximization (EM) algorithm \citep{dempster77a} to
iteratively maximize the likelihood function. We take the latter
approach; for details see the appendix.

\subsection{\Hipparcos\ measurements and their uncertainties}

The sample used in this study is a kinematically unbiased sample of
11,865 nearby main-sequence stars \citep{dehnen98a} from the
\Hipparcos\ catalog \citep{esa97a}, all chosen to have parallaxes
measured at $S/N= \pi/\sigma_{\pi} >10$.  We made no corrections for
Galactic rotation (since this study is simply of stellar velocities
relative to the Sun); indeed we did no processing or correction of the
\Hipparcos\ data beyond making the sample cut.

The ``low dimension'' data $\wwi$ referred to above are the measured
tangential components of each star's ``high dimension'' true
three-dimensional velocity $\vvi$, with
\begin{eqnarray}\displaystyle
\wwi &=& \RRi\,\vvi \nonumber \\
\RRi &=& [\eel\,\eel\T+\eeb\,\eeb\T] \;\;\;,
\end{eqnarray}
where the $\RRi$ are non-square or zero-determinant matrices, and
$\eel$ and $\eeb$ are the tangential unit vectors pointing in the
Galactic latitude and longitude directions for each star.

The $\wwi$ are constructed from the \Hipparcos\ measurements (parallax
and proper motion) as
\begin{equation}
\wwi = \frac{r_0}{\pi_i}\,
       \left[\cos b_i\,\frac{\mathrm{d}l_i}{\mathrm{d}t}\,\eel +
       \frac{\mathrm{d}b_i}{\mathrm{d}t}\,\eeb\right] \;\;\;,
\end{equation}
where $r_0$ is the radius of the Earth's orbit, $\pi_i$ is the star's
parallax, and $l_i$ and $b_i$ are its Galactic latitude and longitude,
and the $\cos b_i$ factor takes into account the spherical geometry.
This calculation ignores the Lutz-Kelker bias \citep{lutz73a}, but
this is small for the sample used here.  Since the \Hipparcos\ catalog
reports proper motions in equatorial rather than Galactic coordinates,
the above requires a rotation depending on each star's angular
position on the sky and the epoch (1991.25) of the catalog positions.

The \Hipparcos\ catalog entries, which can be represented by some
vector $\cci$ for each star, come with single-star uncertainty
covariance matrices $\CCi$.  If we can represent the derivative of the
$\wwi$ with respect to the catalog entries $\cci$ by a matrix $\QQi$
\begin{equation}
\mathrm{d}\wwi = \QQi\,\mathrm{d}\cci \;\;\;,
\end{equation}
then the measurement uncertainty covariances $\SSi$ for the $\wwi$ are
given by
\begin{equation}
\SSi = \QQi\,\CCi\,\QQi\T \;\;\;.
\end{equation}
This is accurate only in the limit of small parallax errors, which is
fine for this sample.  In addition, this whole procedure ignores
star-to-star covariances, which could be significant, but which are
not reported in the \Hipparcos\ catalog.

\section{Results}

\begin{figure}
\includegraphics{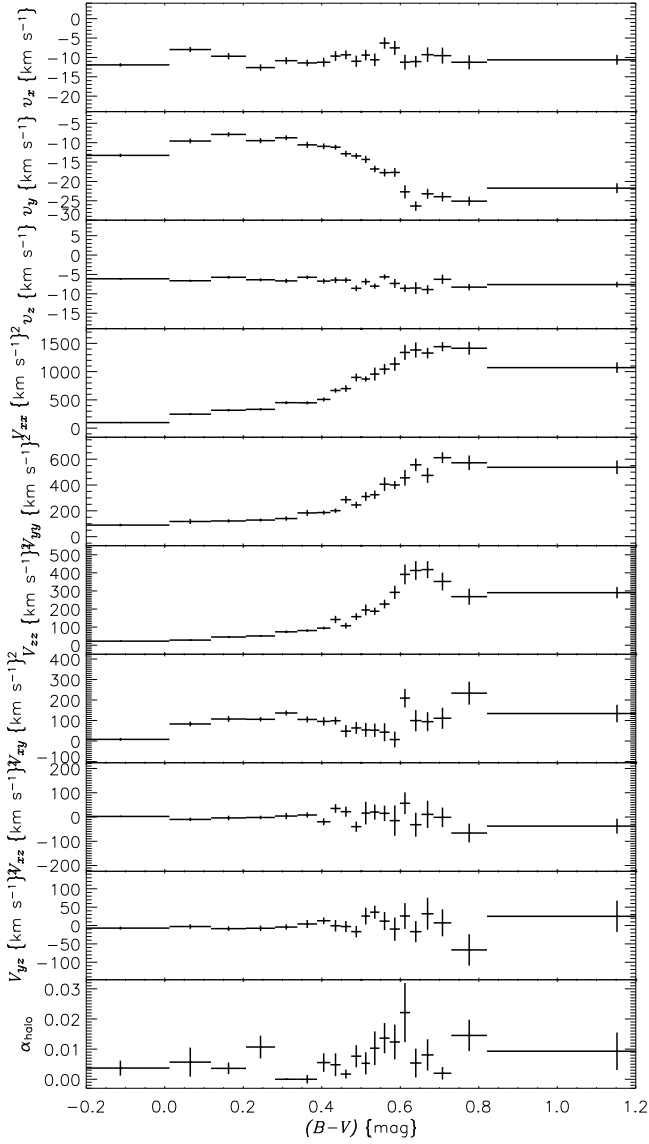}
\caption{The best-fit parameters of the model as a function of stellar
color for the 20 color-selected subsamples.  The off-diagonal elements
of the velocity variance tensor $\VVdisk$ have been scaled by
square-roots of products of the diagonal
elements.\label{fig:lsr_1_parameters}}
\end{figure}

Following the general approach of \cite{dehnen98a}, 20 color-selected
subsamples of stars ($\approx 594$ stars each) were made by cutting
the color-sorted star list into equal-sized pieces (as closely as
possible), and, for each subsample, the 10 parameters $\vvdisk$,
$\VVdisk$, and $\alphahalo$ were found by the optimization described
above.  Figure~\ref{fig:lsr_1_parameters} shows the 10 parameters for
each of the 20 subsamples.  The vertical error-bars on the points
indicate uncertainties computed with 20 independent bootstrap
resamplings of the data in each of the subsamples.  Redder (and
therefore longer-lived) stellar populations have larger velocity
dispersions.

\begin{figure}
\includegraphics{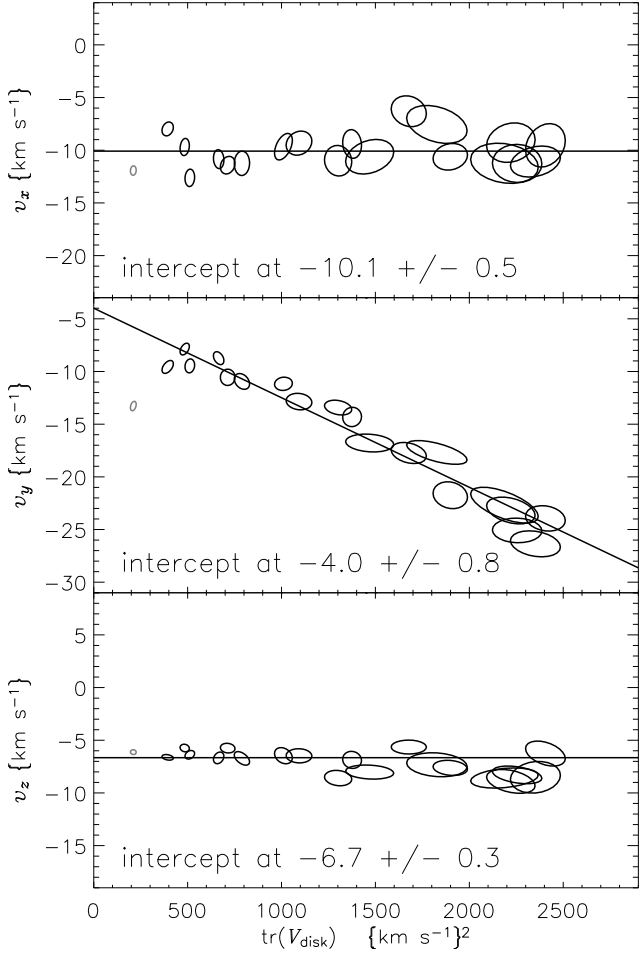}
\caption{The mean velocity $\vvdisk$ as a function of total velocity
variance $\tr(\VVdisk)$ for the determination of the local standard of
rest (LSR).  In each panel, the ellipses indicate the one-sigma
uncertainty regions (from bootstrap resampling---see text) of the
measurements.  The linear fit of $V$ \vs\ $S^2$ was performed with the
projected Gaussian mixtures algorithm because it accounts correctly
for the finite errors in both dimensions (see text).  The point shown
in grey was excluded from the fit because the stars in that subsample
are very short-lived (see text).  The reported uncertatinties on the
intercepts are from 20 bootstrap resamplings of the ellipsoidal points
shown.\label{fig:lsr_1_lsr}}
\end{figure}

Table~\ref{tab:lsr_1_parameters} gives the 10 parameters,
uncertainties, and uncertainty correlation matrix, for one of the
subsamples.  Figure~\ref{fig:lsr_1_lsr} shows the mean $\vvdisk$ of
the disk velocity distribution as a function of the trace
$\tr(\VVdisk)$ of its variance tensor $\VVdisk$.  The mean velocity in
the $\eey$ direction is a strong function of velocity dispersion; this
linear dependence justifies the standard methodology for determination
of the local standard of rest (LSR).

Operationally, the LSR is defined to be the mean velocity for a
hypothetical population of stars with zero velocity distribution, \ie,
the extrapolation to $\tr(\VVdisk) = 0$ of the trend shown in
Figure~\ref{fig:lsr_1_lsr}.  The points in Figure~\ref{fig:lsr_1_lsr}
have significant uncertainties in both dimensions, so fitting a line
responsibly is not trivial.  For this purpose we use again the
projected mixtures of Gaussians procedure described above, but now
there are 19 2-dimensional data points $\wwi$, the $\wwi$ and $\vvi$
are the same (\ie, the $\RRi$ are the identity matrices), and we only
fit a single Gaussian ellipsoid.  The straight line shown in the
$\eey\T\,\vvdisk$ ($v_y$) panel of Figure~\ref{fig:lsr_1_lsr} is the
principal eigenvector of the best-fit Gaussian.  The $\eex\T\,\vvdisk$
and $\eez\T\,\vvdisk$ ($v_x$ and $v_z$) panels show simply weighted
averages.  The errors in the fit are computed by bootstrap resampling
the 19 samples themselves.

The fits shown in Figure~\ref{fig:lsr_1_lsr} provide an intercept
corresponding to the estimated velocity relative to the Sun of a
hypothetical population with vanishing velocity dispersion.  The
velocity of the Sun relative to this LSR is therefore
\begin{equation}
\vvsun = -\vvlsr
\end{equation}
\begin{equation}
\vvsun = [10.1\pm0.5~\mathrm{km\,s^{-1}}]\,\eex
       + [ 4.0\pm0.8~\mathrm{km\,s^{-1}}]\,\eey
       + [ 6.7\pm0.2~\mathrm{km\,s^{-1}}]\,\eez \;\;\;.
\end{equation}

Recall that rejection of halo stars was accomplished by fitting the
velocity field with two Gaussians, one of which was fixed at the halo
parameters given in equation~(\ref{eq:haloparameters}).  Re-fitting
with the halo velocity dispersion increased to
$150~\mathrm{km\,s^{-1}}$ changes the inferred LSR by much less
than the magnitude of its uncertainty.

\begin{figure}
\includegraphics{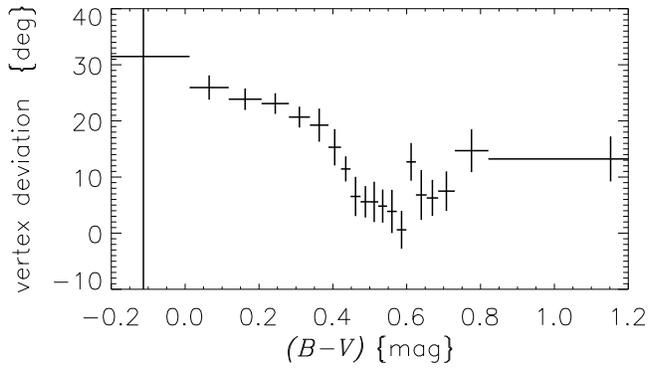}
\caption{The vertex deviation (see text for definition) as a function
of color.\label{fig:lsr_1_vertex}}
\end{figure}

The vertex deviation is defined to be the angle between the $x$ axis
and the projection onto the $x$--$y$ plane of the eigenvector
corresponding to the largest eigenvalue of the velocity variance
tensor $\VVdisk$.  The vertex deviation is shown as a function of
stellar color for the 20 subsamples in Figure~\ref{fig:lsr_1_vertex}.

\section{Algorithm tests}

To test the algorithm, 20 subsamples of artificial 3-dimensional
stellar velocities $\vvi$ were generated with a mixtures of Gaussians
random sample generator.  The distribution was made with two
Gaussians, one for the halo, with parameters as assumed above, and one
for the disk, with mean $\vvdisk$ and variance tensor $\VVdisk$
different for each subsample.  The artificial $\vvi$ were projected
into artificial measurements $\wwi$ using the same $\RRi$ as in the
real 20 subsamples, and errors were added, drawn from two-dimensional
Gaussian ellipsoids with the same variances as the measurement
uncertainty covariance tensors $\SSi$.  \Ie, the artificial data were
given all of the observational properties of the real data (modulo the
assumptions of this study).

For subsamples with mean color $(B-V)< 0.1~\mathrm{mag}$, the
artificial variance tensor $\VVdisk$ was set to the measured value
(shown in Figure~\ref{fig:lsr_1_parameters}) for the subsample with
$(B-V)\approx 0.05~\mathrm{mag}$, and for subsamples with mean $(B-V)>
0.6~\mathrm{mag}$, the artificial tensor $\VVdisk$ was set to the
measured value for the subsample with $(B-V)\approx
0.67~\mathrm{mag}$.  In between, \ie, for artificial subsamples with
mean color $0.1 < (B-V) < 0.6~\mathrm{mag}$, the variance tensor was
made to vary quadratically with color, so as to approximate the
appearance of the true observations (and span the range of observed
variance tensors.  The artificial mean $\vvdisk$ was set to a linear
function of the trace $\tr(\VVdisk)$ of the variance.

\begin{figure}
\includegraphics{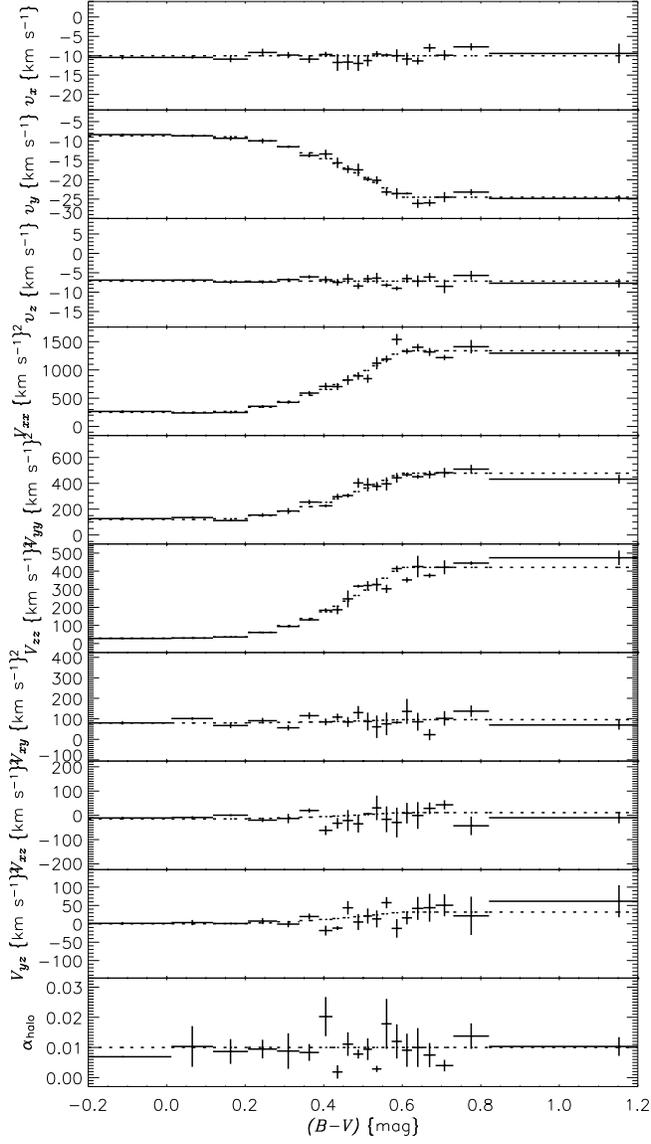}
\caption{Same as Figure~\ref{fig:lsr_1_parameters}, but for the
artificial data (see text).  The dotted lines indicate the input
values used to make the artificial
data.\label{fig:lsr_1fake_parameters}}
\end{figure}

\begin{figure}
\includegraphics{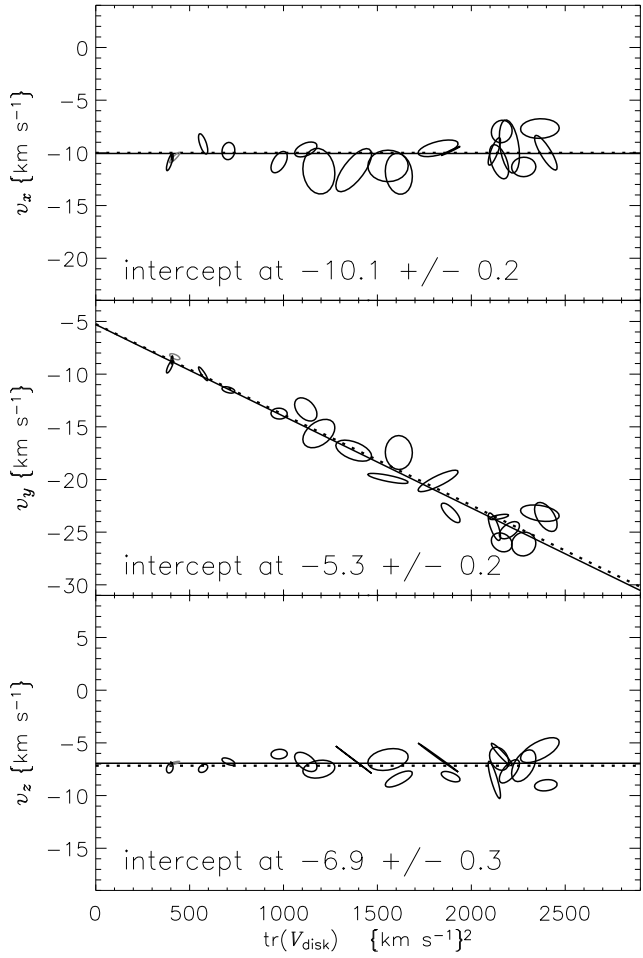}
\caption{Same as Figure~\ref{fig:lsr_1_lsr}, but for the artificial data
(see text).  In each panel there is a dotted line indicating the input
values used to make the artificial data.\label{fig:lsr_1fake_lsr}}
\end{figure}

Exactly the same fitting code and bootstrap analysis was applied to
the artificial data as was applied to the real data.  The results are
shown in Figures~\ref{fig:lsr_1fake_parameters} and
\ref{fig:lsr_1fake_lsr}, along with the input values used to make the
artificial data.  The best-fit parameters are, except for a couple of
samples, within one standard deviation of the input paramters.  More
importantly for present purposes, the LSR is very well determined; the
algorithm returns the correct LSR velocity to well within one standard
deviation.  We conclude that the algorithm is not significantly
biased.

\section{Generalized multi-Gaussian disk}

Perhaps the biggest limitation of LSR measurements like this one is
that the disk velocity distribution function is far from Gaussian; it
is not even unimodal \citep{dehnen98b, skuljan99a, chereul98a}; one of
the primary goals of our future work is to explore the complexities of
disk star velocities.  As a baby step towards checking the influence
of disk velocity non-Gaussianity on the LSR determination, the model
was generalized to allow for not just one Gaussian ellipsoid to fit
the each color subsample's disk velocity distribution function but
$\Kdisk>1$ Gaussians, all constrained to have the same mean.  Models
with $\Kdisk>1$ have the freedom to have larger ``tails'' to the
velocity distribution, and for those tails to be rotated or twisted in
velocity space relative to the core of the velocity distribution.

The generalized model is optimized by an algorithm constructed exactly
parallel to that of the $\Kdisk=1$ model, but now there are
$4+6\,\Kdisk$ free parameters for each subsample.

Increasing $\Kdisk$ increases the goodness-of-fit (of course), but at
very large $\Kdisk$, the data will be ``overfit''.  To determine the
optimal value of $\Kdisk$ for each of the 20 stellar subsamples (the
optimal $\Kdisk$ will, in general, be different for different
subsamples, of course), a ``jackknife likelihood'' was computed: For
each of 5~iterations, a randomly selected 10~percent of each subsample
was removed and put aside as a ``test set.''  Fitting (\ie, parameter
determination by maximum likelihood) was performed on the remaining
90~percent, and the likelihood of the test set was tested within the
context of the best-fit model.  The logarithms of the jackknife
likelihoods for the five iterations were averaged and the $\Kdisk$
with the best jackknife likelihood was chosen for each subsample.

\begin{figure}
\includegraphics{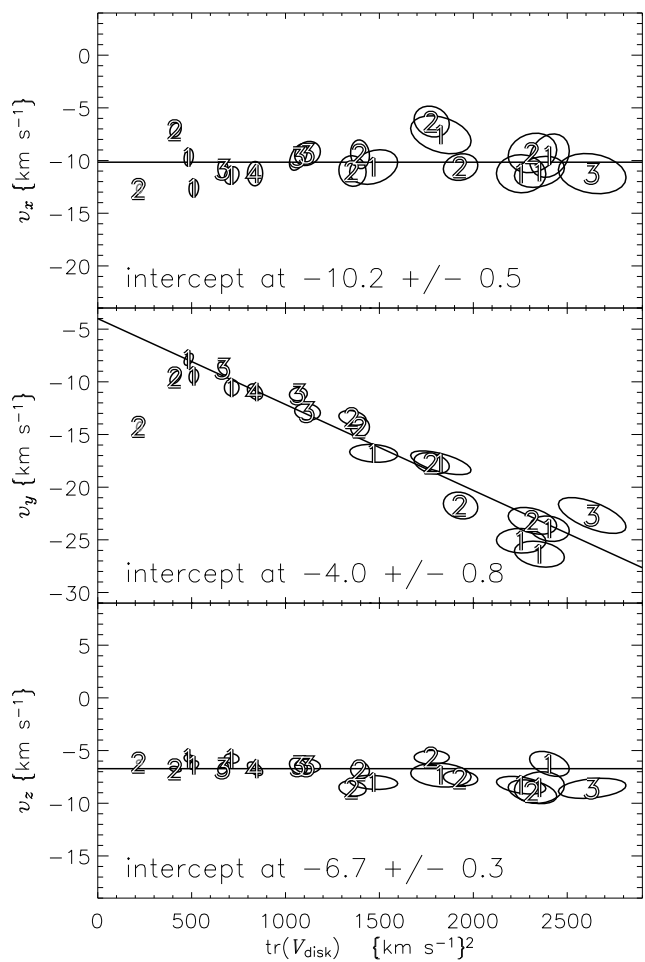}
\caption{Same as Figure~\ref{fig:lsr_1_lsr}, but for the generalized
model in which the disk velocity distribution is fit with a mixture of
$\Kdisk$ Gaussian ellipsoids with common mean.  Each datapoint shows
the result for that subsample for the optimal jackknife value of
$\Kdisk$ (see text), and is labeled by that value of
$\Kdisk$.\label{fig:lsr_bestngauss_lsr}}
\end{figure}

Figure~\ref{fig:lsr_bestngauss_lsr} shows the LSR determination when
the generalized model is used and each sample is fit with the optimal
jackknife likelihood value of $\Kdisk$.  The velocity of the Sun
relative to the LSR we find when using the optimal $\Kdisk$ values is
\begin{equation}
\vvsun = [10.2\pm0.5~\mathrm{km\,s^{-1}}]\,\eex
       + [ 4.0\pm0.8~\mathrm{km\,s^{-1}}]\,\eey
       + [ 6.7\pm0.2~\mathrm{km\,s^{-1}}]\,\eez \;\;\;.
\end{equation}
This is extremely similar to (much closer than 1 standard deviation
away from) that found using $\Kdisk=1$ (Figure~\ref{fig:lsr_1_lsr}).
This suggests that that the assumption of Gaussianity is not strongly
affecting the results.

\section{Discussion}

In the above, we developed and used a novel algorithm to infer the
three-dimensional velocity distribution from a kinematically unbiased
sample of \Hipparcos\ stars.

The local velocity dispersion is a strong function of stellar color,
and the mean velocity of a color-selected stellar population is a
linear function of its velocity variance; this confirms previous
results \citep[\eg, ][]{dehnen98a}.  The extrapolation of this
relation to zero velocity dispersion provides an estimate of the local
standard of rest (LSR), which is found to be
\begin{equation}
\vvsun = [10.1\pm0.5~\mathrm{km\,s^{-1}}]\,\eex
       + [ 4.0\pm0.8~\mathrm{km\,s^{-1}}]\,\eey
       + [ 6.7\pm0.2~\mathrm{km\,s^{-1}}]\,\eez \;\;\;,
\end{equation}
where $\eex$, $\eey$, and $\eez$ are unit vectors pointing in the
directions of the standard Galactic velocity components $U$, $V$, and
$W$.  This result is similar to previouw LSR determinations; we
compare this result to one previous study below.  Our answer did not
change much when we relaxed the assumption that the disk star velocity
distributions can be modeled as Gaussians.

We also showed that it is possible to robustly and reliably solve a
missing data problem in astrophysics: the reconstruction of aspects of
the three-dimensional velocity distribution function from individual
velocity measurements every one of which is missing data in the radial
direction.

\citet{dehnen98a}, using the same subsample of the same data set, find
a somewhat different Solar velocity relative to the LSR; they find
\begin{equation}
\vvsun = [10.00\pm0.36~\mathrm{km\,s^{-1}}]\,\eex
       + [ 5.25\pm0.62~\mathrm{km\,s^{-1}}]\,\eey
       + [ 7.17\pm0.38~\mathrm{km\,s^{-1}}]\,\eez \;\;\;,
\end{equation}
They also find a lower mean velocity variance for the long-lived disk
stars.  These two differences are probably related, since the LSR is
determined by fitting the relationship between velocity and velocity
variance.  Although the studies agree to within about one standard
deviation, better agreement might be expected since both studies are
using identical data subsets of the same data set.  Both studies have
made the incorrect assumption that the stellar velocity distribution
is Gaussian; \citet{dehnen98a} did so in subtracting a measurement
uncertainty variance from the measured velocity variance.  The method
presented here has been shown (using the generalized multi-Gaussian
disk model) to be insensitive to the Gaussianity of the velocity
distribution.  The incorrect assumption of Gaussianity, entering
differently in the \citet{dehnen98a} investigation, may account for
the difference between the results.  Because our method involves the
optimization of a well-defined objective, because we have tested our
method successfully with artificial data, and because we have been
able to relax the Gaussian assumption, we prefer our result.
Certainly the differences show that stellar velocity studies have
become precise enough that algorithms matter.  However, it must be
emphasized that the true velocity distribution is far from a unimodal
Gaussian \citep{dehnen98b, skuljan99a, chereul98a}, so it is not clear
that it is possible to make a ``correct'' LSR determination at all.

\acknowledgments We thank the \Hipparcos\ team for the generous
release of the catalog.  We also thank Walter Dehnen and James Binney
for use of their kinematically unbiased sample and Scott Tremaine and
Matias Zaldarriaga for useful discussions.  This research made use of
the NASA Astrophysics Data System.  MRB and DWH are partially
supported by NASA LTSA grant NAG5-11669 and NSF grant PHY-0101738; KVJ
is partially supported by NASA LTSA grant NAG5-9064 and NSF CAREER
award AST-0133617.  STR is funded in part by the LEARN project of IRIS
and by NSERC and the Canada Research Chairs Program.  KVJ gratefully
acknowledges the hospitality of the NYU Physics Department during the
summer visit when this work was initialized.

\appendix
\section{Fitting Mixtures with Incomplete Data using
the EM algorithm}

The EM algorithm \citep{dempster77a} can be used to optimize the
likelihood function of a probabilistic model involving incomplete
observation data (hidden variables). Starting from user-supplied
starting parameters, its iterations generate a sequence of parameters
which monotonically increase the likelihood of a fixed data set under
the model; thus it finds locally maximum likelihood parameters.

EM proceeds by optimizing, at each point in parameter space, a new
function which is a strict lower bound on the data likelihood.  This
new functions depends on the original model parameters as well as on
some extra auxiliary quantities introduced by EM.  The EM algorithm
iteratively increases the lower bound by coordinate ascent: first (the
``M~step'') the original model parameters are optimized (holding the
auxiliary quantities fixed) and then (the ``E~step'') the auxiliary
quantities are optimized (with the parameters fixed). After the
optimization of the auxiliary quantities, the new function actually
becomes equal to the true model likelihood (the bound becomes tight);
thus at each iteration this true likelihood is nondecreasing.  The M
and E~steps are iterated to convergence, which, here as usually, is
identified by extremely small incremental improvement in the
logarithm of the likelihood per iteration.

In particular, for any auxiliary distribution $q(\vv,j|\ww)$ we can
lower bound the model likelihood $\ln p(\ww)$ by a functional
$F(q)$. (In what follows, we slightly abuse notation by using $j$ both
as an index and as a random variable representing the identity of the
mixture component responsible for generating a particular data point.)

\begin{eqnarray}\displaystyle
\ln p(\ww|\theta)
    &=& \ln \sum_j \int_{\vv} p(\ww,\vv,j|\theta)\,\mathrm{d}\vv\nonumber \\
 &\geq& \sum_j \int_{\vv} q\,\ln \frac{p(\ww,\vv,j|\theta)}{q}\,
                               \mathrm{d}\vv 
        = F(\ww|q,\theta)\nonumber \\
 &\geq& F(\ww|q,\theta)
        = \sum_j \int_{\vv} q(\vv,j|\ww)\,
          \left[ \ln p(\ww,\vv,j|\theta) - \ln q(\vv,j|\ww) \right]\,
          \mathrm{d}\vv\nonumber \\
\ln p(\ww|\theta)
 &\geq& F(\ww|q,\theta) = \left\langle \ln p(\ww,\vv,j|\theta)
                            \right\rangle_{q} + {\cal H}(q) \;\;\;,
\end{eqnarray}
where $\theta$ represents the set of model parameters, and ${\cal H}$
is the entropy of the distribution $q$.

Our strategy is now coordinate maximization of $F$.

In the E~step we maximize $F$ with respect to the auxiliary
distribution $q$. It is easy to show (for example by checking that
it saturates the bound on $F$) that the maximizing distribution $q$ is
the conditional distribution of $\vv$ and $j$ given the observations
$\ww$ and the current parameters:
\begin{equation}
\mbox{\textbf{E~step:}}\;\;\;
q(\vv,j) \leftarrow \mathrm{argmax}_q F(\ww|q,\theta) =  p(\vv,j|\ww,\theta) \;\;\;.
\end{equation}

In the M~step we maximize $F$ with respect to the parameters
$\theta$. This maximization reduces to maximization of the expected
complete log likelihood under the current variational distribution
(since the entropy of $q$ does not depend on $\theta$):
\begin{equation}
\mbox{\textbf{M~step:}}\;\;\;
\theta \leftarrow  \mathrm{argmax}_\theta F(\ww|q,\theta) =
\mathrm{argmax}_\theta \sum_j\int_{\vv} q(\vv,j)\,\ln
                  p(\ww,\vv,j|\theta)\,\mathrm{d}\vv \;\;\;.
\end{equation}

\subsection{EM algorithm for projected mixtures}

For the projected mixtures model the parameters consist of the mixture
component amplitudes $\alpha_j$, means $\mmj$, and variances
$\VVj$. The auxiliaries are posterior distributions for each star $i$:
$q(j|\wwi)$ over mixture components and $q(\vv|j,\wwi)$ over the true
velocity (given the observed projected velocity, assuming it came from
a specific component).  In the terninology of this Appendix, the
parameters of the probability distribution for the true 3-space
velocity $\vv$ of each star, including especially the probabilities
that the star was drawn from each of the gaussian velocity components
$j$, are the ``auxiliary quantities,'' and the parameters
(amplitudes, means, and velocity variance tensors) of the velocity
components $j$ are the ``model parameters.''

First we consider the E~step, in which the auxiliary distributions
$q(\vv,j|\wwi,\theta)$ are optimized. In the case of projected
mixtures, the posterior over $\vv,j$ given $\ww$ and the model
parameters $\theta$ is itself a conditional mixture of Gaussians:
\begin{eqnarray}\displaystyle
\mbox{\textbf{E~step:}}\;\;\;
   p(\vv,j|\ww) &=& p(j|\ww)\,p(\vv|j,\ww)\nonumber \\
       p(j|\wwi) &=& \frac{\alphaj\,\normal(\ww|\RR\,\mmj,\TT_j)}
                     {\displaystyle
                     \sum_k \alphak\,\normal(\ww|\RR\,\mmk,\TT_k)}\nonumber \\
   p(\vv|\wwi,j) &=& \normal(\vv|\bij,\BBij)\nonumber \\
            \bij &=& \mmj+\VVj\,\RRi\T\,\TTij\inv\,(\ww-\RRi\,\mmj)\nonumber \\
           \BBij &=& \VVj-\VVj\,\RRi\T\,\TTij\inv\,\RRi\,\VVj\T \;\;\;;
\end{eqnarray}
Thus, the optimal choice for $q(\vv,j|\wwi,\theta)$ is:
\begin{eqnarray}\displaystyle
   q(\vv,j|\wwi) &=& q(j|\wwi)\,q(\vv|j,\wwi)\nonumber \\
   q(j|\wwi) &=& q_{ij} = p(j|\wwi)\nonumber \\
   q(\vv|\wwi,j) &=& \normal(\vv|\bij,\BBij)
\end{eqnarray}

For the M~step updates, we must explicitly write out the form of the
functional $F$ and take its partial derivatives with respect to each
set of model parameters:

\begin{eqnarray}\displaystyle
F &=& \sum_i  \left\langle \ln p(\wwi|\vvi) + \ln p(\vvi|j) + \ln p(j)
  \right\rangle_{q_i} + {\cal H}(q_i) \nonumber \\
&=& -\frac{1}{2} \sum_i \sum_j q_{ij} \left[
\left\langle
(\wwi - \RRi\vv)\T \SSi\inv (\wwi - \RRi\vv)
(\vv - \mmj)\T \VVj\inv (\vv - \mmj)
\right\rangle_{q(\vv|\wwi,j)}
\rule[-1ex]{0ex}{3ex}\right.\nonumber \\  
&& \quad\quad\quad\quad\quad\quad
\left.\rule[-1ex]{0ex}{3ex}               
+ \ln \det \SSi  + \ln \det \VVj + \log \alpha_i \right] \nonumber \\
&=& -\frac{1}{2} \sum_i \sum_j q_{ij} \left[
\wwi\T\SSi\inv\wwi +\mmj\T\VVj\inv\mmj
- 2\wwi\T\SSi\inv\RRi\bbij -2\mmj\T\VVj\inv\bbij
\rule[-1ex]{0ex}{3ex}\right.\nonumber \\  
&& \quad\quad\quad\quad\quad\quad
\left.\rule[-1ex]{0ex}{3ex}               
+ \mathrm{Trace}\left[ (\RRi\T\SSi\inv\RRi+\VVj\inv)(\bbij\bbij\T +\BBij)\right]
\rule[-1ex]{0ex}{3ex}\right.\nonumber \\  
&& \quad\quad\quad\quad\quad\quad
\left.\rule[-1ex]{0ex}{3ex}               
+ \ln \det \SSi + \ln \det \VVj + \log \alpha_i \right]
\end{eqnarray}

Deriving the E~step and M~step for the particular projected mixtures
model given above leads to the following update equations:

\begin{eqnarray}\displaystyle
\mbox{\textbf{M~step:}}\;\;\;
\alphaj &\leftarrow& \frac{1}{N}\,\sum_i \qij\nonumber \\
   \mmj &\leftarrow& \frac{1}{\qqj}\,\sum_i \qij\,\bij\nonumber \\
   \VVj &\leftarrow& \frac{1}{\qqj}\,\sum_i \qij
                     \left[(\mmj-\bij)\,(\mmj-\bij)\T+\BBij\right]\nonumber \\
  \TTij &\leftarrow& \RRi\,\VVj\,\RRi\T+\SSi \;\;\;.
\end{eqnarray}

Some care must be taken to implement these equations in a numerically
stable way. In particular, care should be taken to avoid underflow
when computing the ratio of a small probability over the sum of other
small probabilities. Notice that we don't have to explicitly enforce
constraints on parameters, \eg, keeping covariances symmetric and
positive definite, since this is taken care of by the updates.  For
example, the update equation for $\VVj$ is guaranteed by its form to
produce a symmetric nonnegative definite matrix.

\clearpage

\clearpage
\begin{deluxetable}{ccc|rrrrrrrrrr}
\rotate
\tablenum{1}
\tablewidth{0pc}
\tablecaption{example parameter set for the subsample with $\subsamplecolor$
  \label{tab:lsr_1_parameters}
}
\tablehead{
  \colhead{parameter} &
  \colhead{value} &
  \colhead{units} &
  \multicolumn{10}{c}{correlation matrix of the (squared) uncertainties}
}
\startdata
$       \eex\T\,\vvdisk$ & $ -9.3\pm  1.9$ &    $\mathrm{km\,s^{-1}}$ &
$ 1.00$ & $ 0.20$ & $ 0.05$ & $ 0.10$ & $-0.14$ & $ 0.38$ & $-0.49$ & $-0.30$ & $ 0.16$ & $-0.53$ \\
$       \eey\T\,\vvdisk$ & $-23.2\pm  1.3$ &    $\mathrm{km\,s^{-1}}$ &
$ 0.20$ & $ 1.00$ & $ 0.39$ & $-0.23$ & $-0.37$ & $-0.08$ & $ 0.12$ & $ 0.18$ & $-0.07$ & $ 0.24$ \\
$       \eez\T\,\vvdisk$ & $ -8.9\pm  1.1$ &    $\mathrm{km\,s^{-1}}$ &
$ 0.05$ & $ 0.39$ & $ 1.00$ & $-0.42$ & $-0.17$ & $ 0.03$ & $ 0.09$ & $ 0.28$ & $-0.30$ & $-0.02$ \\
$ \eex\T\,\VVdisk\,\eex$ & $1329.\pm  95.$ &  $\mathrm{km^2\,s^{-2}}$ &
$ 0.10$ & $-0.23$ & $-0.42$ & $ 1.00$ & $ 0.37$ & $-0.04$ & $-0.10$ & $-0.19$ & $ 0.36$ & $-0.24$ \\
$ \eey\T\,\VVdisk\,\eey$ & $ 474.\pm  58.$ &  $\mathrm{km^2\,s^{-2}}$ &
$-0.14$ & $-0.37$ & $-0.17$ & $ 0.37$ & $ 1.00$ & $-0.30$ & $ 0.21$ & $ 0.07$ & $ 0.53$ & $-0.02$ \\
$ \eez\T\,\VVdisk\,\eez$ & $ 418.\pm  47.$ &  $\mathrm{km^2\,s^{-2}}$ &
$ 0.38$ & $-0.08$ & $ 0.03$ & $-0.04$ & $-0.30$ & $ 1.00$ & $ 0.04$ & $ 0.02$ & $-0.05$ & $-0.47$ \\
$ \eex\T\,\VVdisk\,\eey$ & $  95.\pm  46.$ &  $\mathrm{km^2\,s^{-2}}$ &
$-0.49$ & $ 0.12$ & $ 0.09$ & $-0.10$ & $ 0.21$ & $ 0.04$ & $ 1.00$ & $ 0.27$ & $ 0.00$ & $ 0.15$ \\
$ \eex\T\,\VVdisk\,\eez$ & $  11.\pm  56.$ &  $\mathrm{km^2\,s^{-2}}$ &
$-0.30$ & $ 0.18$ & $ 0.28$ & $-0.19$ & $ 0.07$ & $ 0.02$ & $ 0.27$ & $ 1.00$ & $-0.54$ & $ 0.19$ \\
$ \eey\T\,\VVdisk\,\eez$ & $  32.\pm  44.$ &  $\mathrm{km^2\,s^{-2}}$ &
$ 0.16$ & $-0.07$ & $-0.30$ & $ 0.36$ & $ 0.53$ & $-0.05$ & $ 0.00$ & $-0.54$ & $ 1.00$ & $ 0.05$ \\
$            \alphahalo$ & $0.0081\pm0.0052$ &                          &
$-0.53$ & $ 0.24$ & $-0.02$ & $-0.24$ & $-0.02$ & $-0.47$ & $ 0.15$ & $ 0.19$ & $ 0.05$ & $ 1.00$ \\
\enddata
\end{deluxetable}

\clearpage

\clearpage

\clearpage

\clearpage

\clearpage


\begin{thebibliography}{17}
\expandafter\ifx\csname natexlab\endcsname\relax\def\natexlab#1{#1}\fi

\bibitem[{{Bienaym{\' e}}(1999)}]{bienayme99a}
{Bienaym{\' e}}, O. 1999, \aap, 341, 86

\bibitem[{{Binney} {et~al.}(2000){Binney}, {Dehnen}, \& {Bertelli}}]{binney00a}
{Binney}, J., {Dehnen}, W., \& {Bertelli}, G. 2000, \mnras, 318, 658

\bibitem[{Chereul {et~al.}(1998)Chereul, Cr{\'e}z{\'e}, \&
  Bienaym{\'e}}]{chereul98a}
Chereul, E., Cr{\'e}z{\'e}, M., \& Bienaym{\'e}, O. 1998, \aap, 340, 384

\bibitem[{Chereul {et~al.}(1999)Chereul, Cr{\' e}z{\' e}, \& Bienaym{\'
  e}}]{chereul99a}
Chereul, E., Cr{\' e}z{\' e}, M., \& Bienaym{\' e}, O. 1999, \aaps, 135, 5

\bibitem[{Dehnen(1998)}]{dehnen98b}
Dehnen, W. 1998, \aj, 115, 2384

\bibitem[{{Dehnen}(2000)}]{dehnen00a}
{Dehnen}, W. 2000, \aj, 119, 800

\bibitem[{Dehnen \& Binney(1998)}]{dehnen98a}
Dehnen, W. \& Binney, J.~J. 1998, \mnras, 298, 387

\bibitem[{Dempster {et~al.}(1977)Dempster, Laird, \& Rubin}]{dempster77a}
Dempster, A.~P., Laird, N.~M., \& Rubin, D.~B. 1977, Journal of the Royal
  Statistical Society series B, 39, 1

\bibitem[{{ESA}(1997)}]{esa97a}
{ESA}. 1997, The Hipparcos Catalogue (ESA SP-1136)

\bibitem[{{Fux}(2001)}]{fux01a}
{Fux}, R. 2001, \aap, 373, 511

\bibitem[{Lutz \& Kelker(1973)}]{lutz73a}
Lutz, T.~E. \& Kelker, D.~H. 1973, \pasp, 85, 573

\bibitem[{{Majewski} {et~al.}(2000){Majewski}, {Ostheimer}, {Kunkel}, \&
  {Patterson}}]{majewski00a}
{Majewski}, S.~R., {Ostheimer}, J.~C., {Kunkel}, W.~E., \& {Patterson}, R.~J.
  2000, \aj, 120, 2550

\bibitem[{Quillen(2003)}]{quillen03a}
Quillen, A.~C. 2003, \aj, 125, 785

\bibitem[{{Sirko} {et~al.}(2004){Sirko}, {Goodman}, {Knapp}, {Brinkmann},
  {Ivezi{\' c}}, {Knerr}, {Schlegel}, {Schneider}, \& {York}}]{sirko04a}
{Sirko}, E., {Goodman}, J., {Knapp}, G.~R., {Brinkmann}, J., {Ivezi{\' c}}, {\v
  Z}., {Knerr}, E.~J., {Schlegel}, D., {Schneider}, D.~P., \& {York}, D.~G.
  2004, \aj, 127, 914

\bibitem[{{Skrutskie} {et~al.}(1997){Skrutskie}, {Schneider}, {Stiening},
  {Strom}, {Weinberg}, {Beichman}, {Chester}, {Cutri}, {Lonsdale}, {Elias},
  {Elston}, {Capps}, {Carpenter}, {Huchra}, {Liebert}, {Monet}, {Price}, \&
  {Seitzer}}]{skrutskie97a}
{Skrutskie}, M.~F., {Schneider}, S.~E., {Stiening}, R., {Strom}, S.~E.,
  {Weinberg}, M.~D., {Beichman}, C., {Chester}, T., {Cutri}, R., {Lonsdale},
  C., {Elias}, J., {Elston}, R., {Capps}, R., {Carpenter}, J., {Huchra}, J.,
  {Liebert}, J., {Monet}, D., {Price}, S., \& {Seitzer}, P. 1997, in ASSL Vol.
  210: The Impact of Large Scale Near-IR Sky Surveys, 25--+

\bibitem[{{Skuljan} {et~al.}(1999){Skuljan}, {Hearnshaw}, \&
  {Cottrell}}]{skuljan99a}
{Skuljan}, J., {Hearnshaw}, J.~B., \& {Cottrell}, P.~L. 1999, \mnras, 308, 731

\bibitem[{York {et~al.}(2000)}]{york00a}
York, D. {et~al.} 2000, \aj, 120, 1579

\end{thebibliography}
\end{document}